\documentstyle[12pt,oldlfont]{article}
\normalbaselines
\begin{document}
\bibliographystyle{unsrt}

\vspace{4cm}

\vbox{\vspace{38mm}}
INFN-NA-IV-93/33~~~~~~~~~~~~~~~~~~~~~~~~~~~~~~~~~~~~~~~DSF-T-93/33

\begin{center}
{\LARGE {\bf New relations for two-dimensional Hermite polynomials}}
\end{center}
\vskip5mm
\begin{center}
V.V.Dodonov\\\
{\em Lebedev Physics Institute, Leninsky Prospect 53, 117924 Moscow,
Russia}\\
\vskip3mm
\end{center}
\begin{center}
and V.I.Man'ko\\\
{\em Dipartimento di Scienze Fisiche Universita di Napoli "Federico II"
and I.N.F.N.,Sez.di Napoli}\\\
\end{center}
\begin{center}
{\em and Lebedev Physics Institute, Moscow}\\
\vskip3mm
\end{center}
\begin{abstract}

The effective formulas reducing the two-dimensional Hermite
polynomials to the classical (one-dimensional) orthogonal
polynomials are given.  New one-parameter generating functions
for these polynomials are derived.  Asymptotical formulas for
large values of indices are found.  The applications to the squeezed
one-mode states and to the time-dependent quantum harmonic
oscillator are considered.

\end{abstract}

\newpage

\section{Introduction}

The Hermite polynomials of several variables arise quite naturally
almost in all problems relating to quantum systems described by
means of multidimensional quadratic Hamiltonians:  see, e.g.,  [1-4]
and references therein.  However, until now they were not widely
used by other authors in the papers on quantum mechanics and
quantum optics (for
a few of exceptions see, e.g.,  [5,6]), due to the absence of simple
explicit formulas, which would be convenient for the calculations
and for the analysis of the relations obtained.  The aim of the
present paper is to give the expressions for multidimensional
Hermite polynomials in terms of the well known classical
orthogonal polynomials.  Besides, we shall give new sum rules and
generating functions, as well as asymptotic formulas for various
combinations of the parameters.  Although some of the results
could be found in odd form in other references [1-11], we believe
that bringing them together will be useful for the further
applications.  Just having in mind these applications, we pay a
special attention to the specific sets of parameters defining the
polynomials, which are typical for the quantum mechanical
problems.  We shall consider mainly the case of the Hermite
polynomials of two variables, but some generalizations to higher
dimensions will be also given.

The structure of the paper is as follows.  In the next section we
derive the explicit expressions for the two-dimensional Hermite
polynomials in terms of the Jacobi, Gegenbauer, Legendre, and usual
Hermite polynomials.  The asymptotics of the two-dimensional
Hermite polynomials of zero arguments are considered in Section 3.
New sum rules and generating functions for the ``diagonal''
two-dimensional Hermite polynomials are given in Section 4, and
their generalizations to higher dimensions are considered in Section
5.  Section 6 is devoted to the physical applications.  In that
section we discuss briefly two problems:  the photon statistics in
squeezed mixed (thermal) quantum states, and the transition
probabilities between the energy levels of a quantum oscillator
with a time-dependent frequency.

\section{Relations between two-dimensional Hermite polynomials and
classical orthogonal polynomials of one variable}

The two-dimensional Hermite polynomials $H_{nm}^{\{{\bf R}\}}(y_1
,y_2)$ are
defined by means of the generating function [12],

\begin{equation}\exp[-\frac 12{\bf a}{\bf R}{\bf a}+{\bf a}{\bf R}
{\bf y}]=\sum_{n,m=0}^{\infty}\frac {a_1^na_2^m}{n!m!}H_{nm}^{\{{\bf R}
\}}({\bf y}).\label{1}\end{equation}

Here $a_1$ and $a_2$ are arbitrary complex numbers combined into the
two-dimensional vector ${\bf a}=(a_1,a_2)$ :

\[{\bf a}{\bf R}{\bf a}=\sum_{i,k=0}^2a_iR_{ik}a_k,\qquad {\bf a}
{\bf R}{\bf y}=\sum_{i,k=0}^2a_iR_{ik}y_k,\]

\noindent and ${\bf R}$ is the symmetric matrix

\[{\bf R}=\left(\begin{array}{clcr}
R_{11}&R_{12}\\
R_{12}&R_{22}\end{array}
\right).\]

Let us begin with the case when ${\bf y}=0$ (in the quantum mechanical
applications it corresponds, for example, to the absence of an
external force).   Suppose that $R_{11}\ne 0$ and $R_{22}\ne 0$.
Designating the left-hand side of eq.  (1) as

\[E\equiv\exp\left(-\frac 12R_{11}a_1^2-\frac 12R_{22}a_2^2-R_{12}
a_1a_2\right),\]
we expand it in the following Taylor's series,

\[E=\sum_{p=0}^{\infty}\frac {\left(-R_{11}a_1^2\right)^p}{2^pp!}\left
[1+2\frac {R_{12}a_2}{R_{11}a_1}+\frac {R_{22}a_2^2}{R_{11}a_1^2}\right
]^p.\]
Introducing the notation
\[r=\frac {R_{12}}{\sqrt {R_{11}R_{22}}},\qquad\alpha =\sqrt {\frac {
R_{22}}{R_{11}}}\frac {a_2}{a_1},\]
we can rewrite the square bracket in the right-hand side as

\[\left[1+2r\alpha +\alpha^2\right]=\left[(\alpha +r)^2-(r^2-1)\right
]=(1-r^2)\left[1-(\rho +\gamma )^2\right],\]
where
\[\rho =\frac r{\sqrt {r^2-1}},\qquad\gamma =\frac {\alpha}{\sqrt {
r^2-1}}.\]
Developing the function $\left[1-(\rho +\gamma )^2\right]^p$ with respect
to variable $\gamma$ we arrive at the expansion

\[E=\sum_{p=0}^{\infty}\sum_{k=0}^{2p}\frac {R_{11}^{p-\frac k2}R_{
22}^{\frac k2}}{2^pp!k!}\left(r^2-1\right)^{p-\frac k2}a_1^{2p-k}
a_2^k\frac {{\rm d}^k}{{\rm d}\rho^k}\left(1-\rho^2\right)^p.\]
Taking into account the definition of the {\em associated Legendre
functions\/} [12],

\begin{equation}P_q^s(z)=\frac {(-1)^q}{2^qq!}(z^2-1)^{\frac s2}\frac {
{\rm d}^{q+s}}{{\rm d}z^{q+s}}(1-z^2)^q,\label{2}\end{equation}
we obtain the following formula for the two-dimensional Hermite
polynomial of zero arguments:

\begin{equation}H_{nm}^{\{{\bf R}\}}(0,0)=\mu_{mn}!(-1)^{\frac {m
+n}2}R_{11}^{n/2}R_{22}^{m/2}\left(r^2-1\right)^{\frac {m+n}4}P_{
(m+n)/2}^{|m-n|/2}\left(\frac r{\sqrt {r^2-1}}\right),\label{3}\end{equation}
where
\[\mu_{mn}=\mbox{min}(m,n),\]
and integers $m$, $n$ must have the same parity (otherwise the
right-hand side equals zero).  We draw an attention to the
definition of the associated Legendre functions:  eq.  (2) means
that we consider them in the {\em complex\/} domain, which {\em does not
include\/} the segment $[-1,1]$.  Consequently, the definition adopted in
the present paper differs from the definition used, e.g.,  in the
theory of spherical harmonics and in Refs. [2,3,7], where a special case of
eq. (3) was derived in connection with the calculations of the
transition probabilities between the energy levels of a quantum oscillator
with a time-dependent frequency.  The reason is that in the generic
case we can hardly expect the variable $\rho$ to belong to the segment
$[-1,1]$ (although in some specific but important cases this is possible:
see the last section of the paper).

For coinciding indices we get

\begin{equation}H_{nn}^{\{{\bf R}\}}(0,0)=n!(-\mbox{det}{\bf R})^{\frac
n2}P_n^{}\left(\frac {-R_{12}}{\sqrt {-\mbox{det}{\bf R}}}\right)
,\label{4}\end{equation}
$P_n(z)$ being the usual Legendre polynomial.  Using the relations
between the associated Legendre functions $P_n^m(z)$, the Jacobi
polynomials $P_k^{(\alpha ,\beta )}(z)$, and the Gegenbauer polynomials $
C_q^{\gamma}(z)$ [12,13],

\[P_n^m(z)=\frac {(n+m)!}{2^mn!}\left(z^2-1\right)^{\frac m2}P_{n
-m}^{(m,m)}(z)=\frac {(2m)!}{2^mm!}\left(z^2-1\right)^{\frac m2}C_{
n-m}^{m+\frac 12}(z),\qquad n\ge m,\]
one can write eq. (3) in the following equivalent forms,

\begin{equation}H_{nm}^{\{{\bf R}\}}(0,0)=\frac {m!n!(-1)^{\frac {
m+n}2}}{2^{\frac {|m-n|}2}\left(\frac {m+n}2\right)!}\left[R_{11}^
nR_{22}^m\left(r^2-1\right)^{\mu_{mn}}\right]^{\frac 12}P_{\mu_{m
n}}^{(\frac {|m-n|}2,\frac {|m-n|}2)}\left(\frac r{\sqrt {r^2-1}}\right
),\label{5}\end{equation}

\begin{equation}H_{nm}^{\{{\bf R}\}}(0,0)=\frac {\mu_{nm}!|n-m|!(
-1)^{\frac {m+n}2}}{2^{\frac {|m-n|}2}\left(\frac {|m-n|}2\right)
!}\left[R_{11}^nR_{22}^m\left(r^2-1\right)^{\mu_{mn}}\right]^{\frac
12}C_{\mu_{mn}}^{\frac {|m-n|+1}2}\left(\frac r{\sqrt {r^2-1}}\right
).\label{6}\end{equation}
These expressions hold for arbitrary values of the variable $\rho$.

For nonzero vector {\bf y} function $H_{nm}^{\{{\bf R}\}}(y_1,y_2
)$ can be written as a
finite sum of products of the usual Hermite polynomials [8],

$\left(\frac {R_{11}^nR_{22}^m}{2^{n+m}}\right)^{-\frac 12}$$H_{n
m}^{\{{\bf R}\}}(y_1,y_2)$
\begin{equation}=\sum_{k=o}^{\mu_{mn}}\left(-\frac {2R_{12}}{\sqrt {
R_{11}R_{22}}}\right)^k\frac {n!m!}{(n-k)!(m-k)!k!}H_{n-k}\left(\frac {
\zeta_1}{\sqrt {2R_{11}}}\right)H_{m-k}\left(\frac {\zeta_2}{\sqrt {
2R_{22}}}\right),\label{7}\end{equation}

\noindent where

\begin{equation}\zeta_1=R_{11}y_1+R_{12}y_2,~~~\zeta_2=R_{12}y_1+
R_{22}y_2.\label{8}\end{equation}

 The proof is straightforward. First we express the left-hand side
of eq. (1) as the product of three exponential functions,

\[\exp\left(-\frac 12{\bf a}{\bf R}{\bf a}+{\bf a}{\bf R}{\bf y}\right
)=\exp\left(-z_1^2+2z_1\eta_1\right)\exp\left(-z_2^2+2z_2\eta_2\right
)\exp\left(-\frac {2R_{12}}{\sqrt {R_{11}R_{22}}}z_1z_2\right),\]

\noindent\ where

\[z_1=a_1\sqrt {\frac {R_{11}}2},\qquad z_2=a_2\sqrt {\frac {R_{2
2}}2},\qquad\eta_1=\frac {\zeta_1}{\sqrt {2R_{11}}},\qquad\eta_2=\frac {
\zeta_2}{\sqrt {2R_{22}}}.\]

\noindent\ Then we expand each of three exponentials in the
power series of $z_1$, $z_2$, and $z_1z_2$, respectively, having in mind that
the first two exponentials are the generating functions for the
usual Hermite polynomials.  Combining the terms with the powers
$z_1^{n-k}$ and $z_2^{n-k}$ from the first two expansions with the terms
$(z_1z_2)^k$ arising in the expansion of the third exponential,
$k=0,1,\ldots ,n$, one arrives at formula (7). Note that it remains valid
even when $R_{11}=0$ or $R_{22}=0$, provided one takes the
corresponding limit. For instance, if $R_{22}=0$, then

\begin{equation}\mbox{$H_{nm}^{\{{\bf R}\}}(y_1,y_2)$}=R_{12}^m\sum_{
k=o}^{\mu_{mn}}\left(\frac {R_{11}}2\right)^{\frac {n-k}2}\frac {
(-1)^kn!m!y_1^{m-k}}{(n-k)!(m-k)!k!}H_{n-k}\left(\frac {R_{11}y_1
+R_{12}y_2}{\sqrt {2R_{11}}}\right).\label{9}\end{equation}
Proceeding to the limit $R_{11}=0$ in eq.  (9), we get the formula for
the two-dimensional Hermite polynomial with matrix ${\bf R}=t\sigma_
x$,
where $\sigma_x$ is the standard Pauli matrix,
\[\sigma_x=\left(\begin{array}{clcr}
0&1\\
1&0\end{array}
\right).\]
It reads
\begin{equation}H_{nm}^{\{t\sigma_x\}}(y_1,y_2)=t^{n+m}y_1^my_2^n
\sum_{k=o}^{\mu_{mn}}\frac {n!m!(-y_1y_2t)^{-k}}{(n-k)!(m-k)!k!}.
\label{10}\end{equation}
Taking into account the definition of the associated Laguerre
polynomials [12],

\[L_n^{\alpha}(x)=\sum_{k=0}^n{{n+\alpha}\choose {n-k}}\frac {(-x
)^k}{k!},\]
we can rewrite (10) as follows,

\begin{equation}H_{nm}^{\{t\sigma_x\}}(y_1,y_2)=\mu_{mn}!t^{\nu_{
mn}}(-1)^{\mu_{mn}}y_1^{\frac {(m-n+|m-n|)}2}y_2^{\frac {(n-m+|n-
m|)}2}L_{\mu_{mn}}^{|m-n|}(ty_1y_2),\label{11}\end{equation}
where
\[\nu_{mn}=\mbox{max}(m,n).\]
For coinciding indices we get

\begin{equation}H_{nn}^{\{t\sigma_x\}}(y_1,y_2)=(-1)^nt^nn!L_n(ty_
1y_2).\label{12}\end{equation}

The special cases of eqs. (11) and (12) were obtained earlier in [9,11].

Putting ${\bf y}=0$ in eq. (7) we get a formula for $H_{nm}^{\{{\bf R}
\}}(0,0)$ which may
be the most suitable for the numerical calculations:

\begin{equation}\mbox{$H_{nm}^{\{{\bf R}\}}(0,0)$}=\left(\frac {R_{
11}^nR_{22}^m}{2^{n+m}}\right)^{\frac 12}\sum_{l=0}^{\left[\frac {
\mu_{nm}}2\right]}\frac {(-1)^{\frac {n+m}2}n!m!}{l!\left(\mu_{nm}
-2l\right)!\left(\frac {|n-m|}2+l\right)!}(2r)^{\mu_{nm}-2l},
\label{13}\end{equation}
where $m$ and $n$ must have the same parity. For $m=n$ we have

\begin{equation}\mbox{$H_{nn}^{\{{\bf R}\}}(0,0)$}=\left(\frac {R_{
11}R_{22}}4\right)^{\frac n2}\sum_{l=0}^{\left[\frac n2\right]}\frac {
(-1)^n(n!)^2}{(l!)^2\left(n-2l\right)!}(2r)^{n-2l}.\label{14}\end{equation}

Immediately from the generating function (1) we obtain the
expression for $H_{nm}^{\{{\bf R}\}}(y_1,y_2)$ in terms of $H_{nm}^{
\{{\bf R}\}}(0,0)$ and variables $\zeta_1$, $\zeta_2$
defined by eq. (8):

\begin{equation}H_{nm}^{\{{\bf R}\}}(y_1,y_2)=\sum_{l=0}^n\sum_{k
=0}^m\left(\begin{array}{c}
n\\
l\end{array}
\right)\left(\begin{array}{c}
m\\
k\end{array}
\right)H_{lk}^{\{{\bf
R}\}}(0,0)\zeta_1^{n-l}\zeta_2^{m-k}.\label{15}\end{equation}

\section{Asymptotic formulas for $H_{nm}^{\{{\bf R}\}}(0,0)$}

The asymptotics of the functions $H_{nm}^{\{{\bf R}\}}(0,0)$ for large values
of
indices can be derived, due to eq.  (6), from the asymptotics of the
Gegenbauer polynomials.  Consider first the case when the
argument of these polynomials, $r/\sqrt {r^2-1}$, does not belong to the
interval $[-1,1]$ (remind that $r=R_{12}/\sqrt {R_{11}R_{22}}$ ).  Then the
generalized Laplace-Heine formula (see Theorem 8.21.10 from ref.
[6]) yields

\begin{equation}C_n^{\lambda}(z)\approx\frac {\Gamma (n+\lambda )}{
\Gamma (\lambda )n!}\zeta^n\left(1-\zeta^{-2}\right)^{-\lambda},\qquad
n\gg 1,\qquad\lambda\sim {\cal O}(1),\label{16}\end{equation}
where
\begin{equation}z=\frac 12\left(\zeta +\zeta^{-1}\right),\qquad |
\zeta |>1.\label{17}\end{equation}
The solution of eq. (17) with respect to $\zeta$ reads

\begin{equation}\zeta =z+\epsilon\sqrt {z^2-1}=\frac {r+\epsilon}{\sqrt {
r^2-1}},\qquad\epsilon =\pm 1,\label{18}\end{equation}
where the choice of sign of $\epsilon$ is determined by the requirement
$|\zeta |>1$.  Putting expression (18) into eqs.  (6) and (16) we get for
$n\gg 1$, $|n-m|\sim {\cal O}(1)$,

\begin{equation}H_{nm}^{\{{\bf R}\}}(0,0)\approx\frac {(-1)^{\frac {
n+m}2}\epsilon^{-\frac {|n-m|+1}2}}{\sqrt {2\pi}}\Gamma\left(\frac {
n+m+1}2\right)R_{11}^{n/2}R_{22}^{m/2}\left(r+\epsilon\right)^{\frac {
n+m+1}2}.\label{19}\end{equation}

\noindent\ We have used the identity [12]

\[\Gamma (z)\Gamma (z+\frac 12)=\sqrt {\pi}\Gamma (2z)2^{1-2z}.\]
If parameters $R_{12}$ and $\sqrt {R_{11}R_{22}}$ are real, then $
\epsilon =\mbox{sign}(r)$, and eq.
(19) can be rewritten as follows,

\begin{equation}H_{nm}^{\{{\bf R}\}}(0,0)\approx\frac {(-1)^{\theta_{
nm}}}{\sqrt {2\pi}}\Gamma\left(\frac {n+m+1}2\right)R_{11}^{n/2}R_{
22}^{m/2}\left(\frac {|R_{12}|}{\sqrt {R_{11}R_{22}}}+1\right)^{\frac {
n+m+1}2},\label{20}\end{equation}

\noindent\ where

\[\theta_{nm}=\left\{\begin{array}{cc}
\frac {n-m}2,&\mbox{\quad if}\quad R_{12}<0\\
\frac {n+m}2,&\quad\mbox{if}\quad R_{12}>0\end{array}
\right..\]

However, eqs.  (19) and (20) are not valid for small values of $|
r|$,
since for $R_{12}=0$ and odd values of $n$ and $m$ they do not lead to
the obvious relation $H_{nm}^{\{{\bf R}\}}(0,0)\equiv 0$, which is an
immediate consequence of definition (1).  In this case we may use the
asymptotic formula for the associated Legendre functions given in
ref. [14],

\begin{equation}P_n^{-m}(\cosh\xi )\approx\left(n+\frac 12\right)^{
-m}\left(\frac {\xi}{\sinh\xi}\right)^{1/2}I_m\left(\left[n+\frac
12\right]\xi\right),\label{21}\end{equation}
\[n\gg 1,\qquad m\sim {\cal O}(1),\qquad\mbox{Re}\xi >0,\qquad |\mbox{Im}
\xi |<\pi ,\]
$I_m(z)$ being the modified Bessel function (Bessel function of the
third kind).  Due to eq.  (18),

\[\xi =\log\zeta =\log\left(\frac {r+\epsilon}{\sqrt {r^2-1}}\right
).\]
 Taking into account the relation

\[P_n^m(z)=\frac {(n+m)!}{(n-m)!}P_n^{-m}(z),\]
and a consequence of Stirling's formula,

\begin{equation}\Gamma (x+y)\approx\Gamma (x)x^y,\qquad x\gg 1,\qquad
y\sim {\cal O}(1),\label{22}\end{equation}
we arrive at the expression

\begin{eqnarray}
H_{nm}^{\{{\bf R}\}}(0,0)&\approx&\Gamma ({\cal N}_{nm})(-1)^{\frac {
m+n}2}\left[\epsilon {\cal N}_{nm}R_{11}^nR_{22}^m\left(r^2-1\right
)^{{\cal N}_{nm}}\log\left(\frac {r+\epsilon}{\sqrt {r^2-1}}\right
)\right]^{1/2}\nonumber\\
&\times&I_{|n-m|/2}\left({\cal N}_{nm}\log\left[\frac {r+\epsilon}{\sqrt {
r^2-1}}\right]\right),\end{eqnarray}
where
\[{\cal N}_{nm}=\frac {n+m+1}2.\]

For  real $r>1$ eq. (23) turns into (20) provided one replaces the
modified Bessel function by its asymptotic expression for a large
argument [15],

\[I_a(z)\approx\frac {e^z}{\sqrt {2\pi z}},\qquad |z|\gg 1,\qquad
a\sim {\cal O}(1).\]

Now suppose that $r$ is real, but $|r|<1$. Then

\[\xi =\frac 12\mbox{log}\left(\frac {1+|r|}{1-|r|}\right)-i\frac {
\pi}2\epsilon ,\qquad\epsilon =\mbox{sign}(r),\]
and we may rewrite eq. (23) in terms of the usual Bessel
functions, according to the relation

\[I_k(z)=i^{-k}J_k(iz).\]
It is sufficient to make calculations for $r>0$, having in mind that
the change of sign of $r$ results in an extra factor $(-1)^n=\mbox{$
(-1)^m$}$
before $H_{nm}^{\{{\bf R}\}}(0,0)$, due to eq.  (3).  Thus we get

\begin{eqnarray}
H_{nm}^{\{{\bf R}\}}(0,0)&\approx&(\epsilon i)^{\mu_{nm}}\Gamma (
{\cal N}_{nm})(-1)^{\frac {m+n}2}\left[\frac 12{\cal N}_{nm}R_{11}^
nR_{22}^m\left(1-r^2\right)^{{\cal N}_{nm}}\right]^{1/2}\nonumber\\
&\times&\left[\pi +i\mbox{log}\left(\frac {1+|r|}{1-|r|}\right)\right
]^{1/2}J_{|n-m|/2}\left(\frac 12{\cal N}_{nm}\left[\pi +i\mbox{log}\left
(\frac {1+|r|}{1-|r|}\right)\right]\right).\end{eqnarray}

Taking into account the asymptotics of the Bessel function [12,15],

\[J_k(z)\approx\sqrt {\frac 2{\pi z}}\cos\left(z-\frac 12k\pi -\frac
14\pi\right),\qquad |\mbox{arg}(z)|<\pi ,\qquad |z|\gg 1,\]

we can simplify eq.  (24) as follows,

\begin{eqnarray}
H_{nm}^{\{{\bf R}\}}(0,0)&\approx&\sqrt {\frac 2{\pi}}(\epsilon i
)^{\mu_{nm}}\Gamma ({\cal N}_{nm})(-1)^{\frac {m+n}2}\left[R_{11}^
nR_{22}^m\left(1-r^2\right)^{{\cal N}_{nm}}\right]^{1/2}\nonumber\\
&\times&\cos\left(\frac {\pi}2\mu_{nm}+\frac i2{\cal N}_{nm}\mbox{log}\left
[\frac {1+|r|}{1-|r|}\right]\right)\nonumber\\
\nonumber\\
&\equiv&\frac {(-1)^{\frac {m+n}2}}{\epsilon^n\sqrt {2\pi}}\Gamma
({\cal N}_{nm})\left[R_{11}^nR_{22}^m\right]^{1/2}\left[\left(1+|
r|\right)^{{\cal N}_{nm}}+(-1)^n\left(1-|r|\right)^{{\cal N}_{nm}}\right
].\end{eqnarray}

If $r=0$, this expression exactly equals zero for odd values of $
n$
and $m$ (remind that $m$ and $n$ must have the same parity).
Moreover, it coincides with eq. (20) provided
$\left(1+|r|\right)^{{\cal N}_{nm}}\gg\left(1-|r|\right)^{{\cal N}_{
nm}}$.

The exceptional case when coefficient $\rho$ (the argument of the
Gegenbauer or Jacobi polynomials) belongs to the interval $[-1,1]$, so
that $\rho =\cos\theta$, corresponds to $r=i\cot\theta$. Then we use
the Hilb--like asymptotics of the Jacobi polynomial, given by
Theorem 8.21.12 of Ref. [13],

\[P_q^{(\alpha ,\beta )}(\cos\theta )\approx\frac {\Gamma (q+\alpha
+1)}{N^{\alpha}n!}\left(\sin\frac {\theta}2\right)^{-\alpha}\left
(\cos\frac {\theta}2\right)^{-\beta}\left(\frac {\theta}{\sin\theta}\right
)^{\frac 12}J_{\alpha}(N\theta ),\]
where $N=n+(\alpha +\beta +1)/2$, $\alpha >-1$, and $\beta$ may be
arbitrary number.
Putting this expression into eq. (5) we arrive at the formula

\begin{equation}H_{nm}^{\{{\bf R}\}}(0,0)\approx\frac {\nu_{nm}!\left
[(-1)^{\mu_{nm}}\theta R_{11}^nR_{22}^m\right]^{1/2}}{(-1)^{\frac {
n+m}2}{\cal N}_{nm}^{\frac {|n-m|}2}(\sin\theta )^{{\cal N}_{nm}}}
J_{\frac {|n-m|}2}\left({\cal N}_{nm}\theta\right),\label{26}\end{equation}
which is valid for ${\cal N}_{nm}\gg 1$, $|n-m|\sim {\cal O}(1)$, $
0\le\theta <\pi$.

Eq. (6) enables also to find the asyptotics of $H_{nm}^{\{{\bf R}
\}}(0,0)$, when
$|m-n|\gg 1,$ $\mu_{nm}\sim {\cal O}(1)$. The definition of the
Gegenbauer polynomial [12],

\[C_q^{\lambda}(z)=\sum_{k=0}^{\left[\frac q2\right]}\frac {(-1)^
k\Gamma (\lambda +q-k)}{\Gamma (\lambda )k!(q-2k)!}(2z)^{q-2k},\]
with account of eq. (22), yields for $\lambda\gg 1$

\[C_q^{\lambda}(z)\approx\sum_{k=0}^{\left[\frac q2\right]}\frac {
(-1)^k\lambda^{q-k}}{k!(q-2k)!}(2z)^{q-2k}.\]
Comparing this expression with the definition of the Hermite
polynomial [12],

\[H_q(z)=q!\sum_{k=0}^{\left[\frac q2\right]}\frac {(-1)^k(2z)^{q
-2k}}{k!(q-2k)!},\]
we get the asymptotic formula for the Gegenbauer polynomial,

\begin{equation}C_q^{\lambda}(z)\approx\frac {\lambda^{q/2}}{q!}H_
q(z\sqrt {\lambda}),\qquad\lambda\gg 1,\qquad q\sim {\cal O}(1),
\label{27}\end{equation}
and its consequence for the two-dimensional Hermite polynomial of
zero arguments,

\begin{eqnarray}
H_{nm}^{\{{\bf R}\}}(0,0)&\approx&\frac {(-1)^{\frac {m+n}2}}{\sqrt {
\pi}}2^{\frac {|n-m|}2}\Gamma\left(\frac {\nu_{nm}+1}2\right)\left
[R_{11}^nR_{22}^m\left(r^2-1\right)^{\mu_{nm}}\right]^{1/2}\nonumber\\
&\times&H_{\mu_{mn}}\left(\frac {r\sqrt {|n-m|}}{\sqrt {2(r^2-1)}}\right
),\qquad |n-m|\gg 1,\qquad\mu_{nm}\sim {\cal O}(1).\end{eqnarray}
(We have used again eq. (22)). The limit cases of formula (28) for
small and large values of the argument of the Hermite polynomial
are quite clear.

\section{One-parameter generating function and sum rules for the
``diagonal''polynomials}

Generating function (1) has two auxiliary parameters, $a_1$ and $
a_2$.
However, for the ``diagonal'' polynomials, with $m=n$, a
one-parameter generating function exists.  To find it we choose
vector ${\bf a}$ as follows,

\[a_1=\sqrt {\lambda}\beta^{*},~~~a_2=\sqrt {\lambda}\beta ,\]

\noindent where $\lambda$ is a new complex parameter.  Then multiplying
both sides of eq.  (1) by the factor $\exp(-\beta\beta^{*})$ and
calculating the gaussian integral
\begin{eqnarray*}
G(\lambda )=\frac i{2\pi}\int\exp(-\frac 12{\bf a}{\bf R}{\bf a}+
{\bf a}{\bf f}-\beta\beta^{*})d\beta d\beta^{*}\\
=\sum_{n,m=0}^{\infty}\lambda^m\frac {H_{nm}^{\{{\bf R}\}}({\bf R}^{
-1}{\bf f})}{n!m!}\frac i{2\pi}\int\beta^{*n}\beta^me^{-\beta\beta^{
*}}d\beta d\beta^{*},\end{eqnarray*}

\noindent with the account of the relation

\[\frac i{2\pi}\int\beta^{*n}\beta^me^{-\beta\beta^{*}}d\beta d\beta^{
*}=n!\delta_{nm},\]

\noindent we arrive at the formula ({\bf I} means the unity matrix)

\begin{equation}G(\lambda )\equiv\sum_{n=0}^{\infty}\frac {\lambda^
n}{n!}H_{nn}^{\{{\bf R}\}}({\bf R}^{-1}{\bf f})=\left[\det\left(\lambda
\sigma_x{\bf R}+{\bf I}\right)\right]^{-\frac 12}\exp\left[\frac {
\lambda}2{\bf f}\left(\lambda\sigma_x{\bf R}+{\bf I}\right)^{-1}\sigma_
x{\bf f}\right].\label{29}\end{equation}

  The explicit form of the right-hand side of eq.  (29) reads
(${\bf f}\equiv (f_1,f_2)$)

\begin{equation}G(\lambda )=\left[1+2\lambda R_{12}-\lambda^2\det
{\bf R}\right]^{-\frac 12}\exp\left\{\frac {2\lambda f_1f_2-\lambda^
2\left(f_1^2R_{22}+f_2^2R_{11}-2f_1f_2R_{12}\right)}{2\left[1+2\lambda
R_{12}-\lambda^2\det {\bf R}\right]}\right\}.\label{30}\end{equation}
If ${\bf f}=0$, then this function is nothing but the generating function
of the Legendre polynomials [12], and we again arrive at eq. (4).

Eq.  (30) enables us to represent the ``diagonal'' two-dimensional
Hermite polynomials in the form of a finite sum over the products
of the Laguerre polynomials.  For this purpose we rewrite the
right-hand side of eq.  (30) as follows,

\[G(\lambda )=\left[\left(1-\frac {\lambda}{\lambda_1}\right)\left
(1-\frac {\lambda}{\lambda_2}\right)\right]^{-\frac 12}\exp\left[\frac {
\lambda x_1}{\lambda -\lambda_1}+\frac {\lambda x_2}{\lambda -\lambda_
2}\right],\]

\noindent where
\[\lambda_1=(\sqrt {R_{11}R_{22}}-R_{12})^{-1},\qquad\lambda_2=-(\sqrt {
R_{11}R_{22}}+R_{12})^{-1},\]
\[x_1=\frac {f_1^2R_{22}+f_2^2R_{11}-2\sqrt {R_{11}R_{22}}f_1f_2}{
4\sqrt {R_{11}R_{22}}\left(\sqrt {R_{11}R_{22}}-R_{12}\right)},\]
\[x_2=\frac {f_1^2R_{22}+f_2^2R_{11}+2\sqrt {R_{11}R_{22}}f_1f_2}{
4\sqrt {R_{11}R_{22}}\left(\sqrt {R_{11}R_{22}}+R_{12}\right)}.\]

\noindent Using the known relation [12] for the Laguerre polynomials
\[\frac 1{\sqrt {1-z}}\exp\frac {xz}{z-1}=\sum_{n=0}^{\infty}z^nL_
n^{-\frac 12}(x),\]

\noindent we see that the generating function $G(\lambda )$ can be
represented as
\[G(\lambda )=\sum_{s=0}^{\infty}\left(\frac {\lambda}{\lambda_1}\right
)^sL_s^{-\frac 12}(x_1)\sum_{k=o}^{\infty}\left(\frac {\lambda}{\lambda_
2}\right)^kL_k^{-\frac 12}(x_2).\]

\noindent Thus we obtain the formula
\begin{equation}H_{nn}^{\{{\bf R}\}}({\bf R}^{-1}{\bf f})=(-1)^nn
!\sum_{s=0}^n(R_{12}-\sqrt {R_{11}R_{22}})^s(R_{12}+\sqrt {R_{11}
R_{22}})^{n-s}L_s^{-\frac 12}(x_1)L_{n-s}^{-\frac 12}(x_2).
\label{31}\end{equation}

\noindent\ Assuming ${\bf f}={\bf R}{\bf y}$, we get quite the same
decomposition (31) for the function $H_{nn}^{\{{\bf R}\}}({\bf y})$,
but with the variables $x_1$ and $x_2$
expressed directly in terms of $y_1$ and $y_2$ as follows,

\[x_1=\frac 14\left(1-\frac {R_{12}}{\sqrt {R_{11}R_{22}}}\right)\left
(y_1^2R_{11}+y_2^2R_{22}-2\sqrt {R_{11}R_{22}}y_1y_2\right),\]
\[x_2=\frac 14\left(1+\frac {R_{12}}{\sqrt {R_{11}R_{22}}}\right)\left
(y_1^2R_{11}+y_2^2R_{22}+2\sqrt {R_{11}R_{22}}y_1y_2\right).\]

Formulas like (30) and (31) for the diagonal matrix elements of the
Gaussian density matrix in the Fock basis were obtained also in Ref. [16].

Differentiating the generating function $G(\lambda )$ {\em k\/} times
with respect to $\lambda$ and putting $\lambda =1$ in the final
expression, one can calculate the cumulants $<n(n-1)\cdots (n-k+1)>$,
defined according to the relation

\[\sum_{n=0}^{\infty}\mbox{$n(n-1)\cdots (n-k+1)$}\frac {H_{nn}^{
\{{\bf R}\}}({\bf y})}{n!}=\mbox{$<n(n-1)\cdots (n-k+1)>$}\sum_{n
=0}^{\infty}\frac {H_{nn}^{\{{\bf R}\}}({\bf y})}{n!}.\]

\noindent The first two derivatives yield the ``average value'' $
<n>$
and the "dispersion"

\[\sigma_n=<n^2>-<n>^2=<n(n-1)>+<n>-<n>^2.\]

\noindent Differentiating the preexponential factor in eq.  (29) with
the aid of the matrix formula

\[\frac d{dt}\log\det {\bf A}(t)={\rm T}{\rm r}\left({\bf A}^{-1}\frac {
d{\bf A}}{dt}\right),\]

\noindent we get the following general concise relations,

\begin{equation}<n>({\bf R},{\bf y})=\frac 12{\rm T}{\rm r}{\bf S}
-1+\frac 12{\bf y}{\bf R}{\bf S}^2\sigma_x{\bf R}{\bf y},
\label{32}\end{equation}
\begin{equation}\sigma_n({\bf R},{\bf y})=\frac 12{\rm T}{\rm r}[
{\bf S}({\bf S}-{\bf I})]+\frac 12{\bf y}{\bf R}(2{\bf S}-{\bf I}
){\bf S}^2\sigma_x{\bf R}{\bf y},\label{33}\end{equation}

\noindent where matrix ${\bf S}$ is defined as follows,

\[{\bf S}=(\sigma_x{\bf R}+{\bf I})^{-1}.\]
Of course, these formulas hold provided the series converge.  The
simplest special case corresponds to the matrix ${\bf R}=t\sigma_
x$, when

\[<n>=-\frac t{1+t}+y_1y_2\left(\frac t{1+t}\right)^2,\qquad\sigma_
n=-\frac t{(1+t)^2}+y_1y_2\frac {t^2(1-t)}{(1+t)^3}.\]
These relations have sense for $-1<t<0$.

\section{Generalizations to higher dimensions}

To generalize the results of the previous section to the ``diagonal''
Hermite polynomials of $2N$ variables we introduce $N$ complex
parameters $\lambda_i$, $i=1,2,\ldots ,N$, and choose the auxiliary
parameters $a_j$ as follows,

\[a_i=\sqrt {\lambda_i}\beta_i^{*},~~~a_{i+N}=\sqrt {\lambda_i}\beta_
i,\qquad\mbox{$i=1,2,\ldots ,N$}.\]

\noindent Then the same scheme results in the following expansion,

\begin{eqnarray}
&&\sum_{n_1=0}^{\infty}\ldots\sum_{n_N=0}^{\infty}\frac {\lambda_
1^{n_1}}{n_1!}\frac {\lambda_2^{n_2}}{n_2!}\ldots\frac {\lambda_N^{
n_N}}{n_N!}H_{n_1n_2\ldots n_Nn_1n_2\ldots n_N}^{\{{\bf R}\}}({\bf y}
)\nonumber\\
&=&\left[\det\left(\Lambda\Sigma_x{\bf R}+{\bf I}_{2N}\right)\right
]^{-\frac 12}\exp\left[\frac 12{\bf y}{\bf R}\left(\Lambda\Sigma_
x{\bf R}+{\bf I}_{2N}\right)^{-1}\Sigma_x\Lambda {\bf R}{\bf y}\right
].\end{eqnarray}
\noindent Here ${\bf y}=(y_1,y_2,...y_{2N})$, the $2N$x$2N$ matrix $
\Sigma_x$ is the
$2N$-dimensional analog of the Pauli matrix $\sigma_x$,

\[\Sigma_x=\left(\begin{array}{cc}
0&{\bf I}_N\\
{\bf I}_N&0\end{array}
\right),\]

\noindent${\bf I}$$_N$ is $N\times N$ unit matrix, and the diagonal $
2N$x$2N$- matrix $\Lambda$
reads
\[\Lambda =\sum_{j=1}^N\lambda_j\Lambda_j,\]
where each matrix $\Lambda_j$ has only two nonzero elements:
\[(\Lambda_j)_{jj}=(\Lambda_j)_{j+N,j+N}=1.\]
Note that matrix $\Lambda$ commutes with matrix $\Sigma_x$ but does
not commute, in general, with matrix ${\bf R}$.

Generating function (34) enables to calculate the ``average values''
$<n_1^{\alpha_1}n_2^{\alpha_2}\cdots n_N^{\alpha_N}>$, defined as follows,

\begin{eqnarray*}
\sum_{n_1=0}^{\infty}\ldots\sum_{n_N=0}^{\infty}n_1^{\alpha_1}n_2^{
\alpha_2}\cdots n_N^{\alpha_N}\frac {H_{n_1n_2\ldots n_Nn_1n_2\ldots
n_N}^{\{{\bf R}\}}({\bf y})}{n_1!n_2!\cdots n_{_{}N}!}\nonumber\\
=\mbox{$<n_1^{\alpha_1}n_2^{\alpha_2}\cdots n_N^{\alpha_N}>$}\sum_{
n_1=0}^{\infty}\ldots\sum_{n_N=0}^{\infty}\frac {H_{n_1n_2\ldots
n_Nn_1n_2\ldots n_N}^{\{{\bf R}\}}({\bf y})}{n_1!n_2!\cdots n_{_{}
N}!}.\end{eqnarray*}

\noindent Taking into account the relations

\[\frac {\partial\Lambda}{\partial\lambda_j}=\Lambda_j,\qquad\Lambda_
j\Lambda_k={\bf 0},\quad j\neq k,\qquad\Lambda_j^2=\Lambda_j,\qquad
\Lambda_j\Sigma_x=\Sigma_x\Lambda_j,\]
we get the following formulas,

\begin{equation}<n_j>=\frac 12{\rm T}{\rm r}\left[\Lambda_j{\bf S}\right
]-1+\frac 12{\bf y}{\bf R}\mbox{${\bf S}\Lambda_j{\bf S}\Sigma_x{\bf R}
{\bf y}$},\label{35}\end{equation}
\begin{equation}\sigma_{n_jn_k}=\frac 12{\rm T}{\rm r}\left[\Lambda_
j{\bf S}\Lambda_k{\bf S}\right]+\frac 12{\bf y}{\bf R}{\bf S}\Lambda_
j{\bf S}\Lambda_k{\bf S}\Sigma_x{\bf R}{\bf y},\qquad j\neq k
\label{36}\end{equation}
\begin{equation}\sigma_{n_jn_j}=\frac 12{\rm T}{\rm r}\left[\Lambda_
j{\bf S}\left(\Lambda_j{\bf S}-{\bf I}_{2N}\right)\right]+\frac 1
2{\bf y}{\bf R}\left(2{\bf S}\Lambda_j-{\bf I}_{2N}\right){\bf S}
\Lambda_j{\bf S}\Sigma_x{\bf R}{\bf y},\label{37}\end{equation}

\noindent where
\[\sigma_{n_jn_k}=<n_jn_k>-<n_j><n_k>,\]
and matrix ${\bf S}$ reads
\[{\bf S}=(\Sigma_x{\bf R}+{\bf I}_{2N})^{-1}.\]

\section{Physical applications}

{\bf 1}. Suppose we have some {\em Gaussian distribution function}

\begin{equation}{\cal W}({\bf q})=\left[\det(2\pi {\bf M})\right]^{
-\frac 12}\exp\left\{-\frac 12({\bf q}-\langle {\bf q}\rangle ){\bf M}^{
-1}({\bf q}-\langle {\bf q}\rangle )\right\},\label{38}\end{equation}
where {\bf q} is an $N$-dimensional vector, and {\bf M} is a symmetric
positively definite $N\times N$-matrix, which is nothing but the
covariance matrix,

\[{\bf M}=\left\Vert {\cal M}_{ij}\right\Vert ,\qquad {\cal M}_{i
j}=\langle q_iq_j\rangle -\langle q_i\rangle\langle q_j\rangle ,\]
\[\langle f({\bf q})\rangle\equiv\int f({\bf q}){\cal W}({\bf q})\mbox{d}^
N{\bf q},\qquad\int {\cal W}({\bf q})\mbox{d}^N{\bf q}=1.\]

In this case the {\em characteristic function} $\chi ({\bf a})\equiv
\langle\exp({\bf a}{\bf q})\rangle$ is also
Gaussian,

\begin{equation}\chi ({\bf a})=\exp\left(\frac 12{\bf a}{\bf M}{\bf a}
+{\bf a}\langle {\bf q}\rangle\right).\label{39}\end{equation}
On the other hand, evidently,

\[\chi ({\bf a})=\sum_{n_1,\ldots n_N=0}^{\infty}\frac {a_1^{n_1}
\cdots a_N^{n_N}}{n_1!\cdots n_N!}\langle q_1^{n_1}\cdots q_N^{n_
N}\rangle .\]
Comparing these formulas with eq. (1) we arrive at the relation

\begin{equation}\langle q_1^{n_1}\cdots q_N^{n_N}\rangle =H_{n_1\ldots
n_N}^{\{-{\bf M}\}}\left(-{\bf M}^{-1}\langle {\bf q}\rangle\right
).\label{40}\end{equation}

{\bf 2}.  If $N=2$, ${\bf q}=(p,q)$, and ${\cal W}({\bf q})$ is the
{\em Wigner function\/} of some
one-dimensional quantum system with quadrature components $p$, $q$,
then the probability to observe $n$ quanta in the state described by
${\cal W}({\bf q})$ (in general, it is a {\em mixed squeezed quantum state\/})
is given by the relation [3,5,6,9,17]

\begin{equation}{\cal P}_n={\cal P}_0\frac {H_{nn}^{\{{\bf R}\}}(
y_1,y_2)}{n!},\label{41}\end{equation}
with the following elements of matrix {\bf R},

\begin{equation}R_{11}=R_{22}^{*}=\frac {2\left(\sigma_{pp}-\sigma_{
qq}-2i\sigma_{pq}\right)}{1+2T+4d},\qquad R_{12}=\frac {1-4d}{1+2
T+4d},\label{42}\end{equation}
and vector {\bf y},

\begin{equation}y_1=y_2^{*}=\frac {2\left[\left(2\sigma_{pp}-1+2i
\sigma_{pq}\right)\langle q\rangle +i\left(1-2\sigma_{qq}+2i\sigma_{
pq}\right)\langle p\rangle\right]}{2T-4d-1}.\label{43}\end{equation}

\noindent Here
\begin{equation}d=\det {\bf M}\ge\frac 14,\qquad T=\mbox{Tr}{\bf M}
\ge 1\label{44}\end{equation}
(the inequalities are the consequences of the {\em generalized
uncertainty relations\/} [3], which ensure the positive definiteness of
the statistical operator corresponding to function ${\cal W}({\bf q}
))$.

The probability to have no photons ${\cal P}_0$ is given by the formula
\begin{equation}{\cal P}_0=(d+\frac 12T+\frac 14)^{-\frac 12}\exp\left
[-\frac {\langle p\rangle^2(2\sigma_{qq}+1)+\langle q\rangle^2(2\sigma_{
pp}+1)-4\sigma_{pq}\langle p\rangle\langle q\rangle}{1+2T+4d}\right
].\label{45}\end{equation}

Various special cases of this photon distribution function were investigated
also in [16,18].
Confining ourselves to the case of $\langle p\rangle =\langle q\rangle
=0$ and taking into
account eq. (4), we can simplify (41) as follows,

\begin{equation}{\cal P}_n=\left(d+\frac T2+\frac 14\right)^{-\frac
12}\left(\frac {4d+1-2T}{4d+1+2T}\right)^{\frac n2}P_n\left(\frac {
4d-1}{\left[\left(4d+1\right)^2-4T^2\right]^{\frac 12}}\right).
\label{46}\end{equation}
For {\em pure quantum states}, when $d=1/4$, we have an {\em oscillating \/}
photon distribution function [5,19,20]:

\begin{equation}{\cal P}_{2m}^{(pure)}=\left(\frac 2{1+T}\right)^{\frac
12}\frac {(2m)!}{\left(2^mm!\right)^2}\left(\frac {T-1}{T+1}\right
)^m,\qquad {\cal P}_{2m+1}^{(pure)}=0,\label{47}\end{equation}
To see the ``smoothing'' of these oscillations for {\em mixed\/} quantum
states we use eq. (25). Since in the case under study
\[r=\frac {1-4d}{2\sqrt {T^2-4d}}\le 0,\]
we choose the values of $T$ and $d$ in the domain
\[T-\frac 12>2d\ge\frac 12,\]
to ensure the inequality $|r|<1$. Then we have for $n\gg 1$

\begin{equation}{\cal P}_n\approx\frac {\left[2\sqrt {T^2-4d}+4d-
1\right]^{n+\frac 12}+(-1)^n\left[2\sqrt {T^2-4d}-4d+1\right]^{n+\frac
12}}{\sqrt {\pi n}(1+2T+4d)^{n+\frac 12}\left(T^2-4d\right)^{\frac
14}}.\label{48}\end{equation}

For $d=1/4$ this expression coincides with the Stirling asymptotics
of formula (47). Eq. (48) shows that the oscillations of the photon
distribution function disappear provided

\[n\frac {4d-1}{2\sqrt {T^2-4d}}\geq 1.\]
For highly squeezed mixed states, when $T^2\gg 4d\gg 1$, this criterion
assumes the form $2nd/T\ge 1$.

{\bf 3}. For a quantum harmonic oscillator with a time-dependent
frequency the transition probability from the $n$-th to the $m$-th
stationary level can be expressed as [1-3,7,8]

\begin{equation}W_n^m=\frac {W_0^0}{n!m!}\left|H_{nm}^{\{{\bf R}\}}
(x_1,x_2)\right|^2,\label{49}\end{equation}
with matrix {\bf R} of the form
\[{\bf R}=-\frac 1{\xi}\left(\begin{array}{cc}
-\eta&1\\
1&\eta^{*}\end{array}
\right),\qquad |\xi |^2-|\eta |^2=1,\]
and
\[W_0^0=|\xi |^{-1}.\]
The explicit form of the parameters $\xi$ and $\eta$ is determined by
the solutions of the classical equation

\[\ddot{\epsilon }+\omega^2(t)\epsilon =0,\]
moreover, the ratio $|\eta /\xi |^2$ may be interpreted as a reflection
coefficient from some ``effective potential barrier'' $\omega^2(t
)$, provided
a ``transmitted wave'' is $\exp(i\omega t)$ for $t\to -\infty$.
The parameters $x_1$,
$x_2$ are expressed through the convolution of the function $\epsilon
(t)$ with
the external time-dependent force.  If this force is absent,
$x_1=x_2=0$.  Confining ourselves with this simplest case we can
rewrite eq.  (49), due to eqs.  (5) and (6), in the following
equivalent forms (remind that $m$ and $n$ must be either both even
or both odd):

\begin{equation}W_n^m=\frac {n!m!}{\left[\left(\frac {n+m}2\right
)!\right]^2|\xi |}\left|\frac {\eta}{2\xi}\right|^{|n-m|}\left[P_{
\mu_{nm}}^{\left(\frac {|n-m|}2,\frac {|n-m|}2\right)}\left(\frac
1{|\xi |}\right)\right]^2,\label{50}\end{equation}
\begin{equation}W_n^m=\frac {\mu_{nm}!}{\nu_{nm}!}\left[\frac {|n
-m|!}{\left(\frac {|n-m|}2\right)!}\right]^2\frac 1{|\xi |}\left|\frac {
\eta}{2\xi}\right|^{|n-m|}\left[C_{\mu_{nm}}^{\frac {|n-m|+1}2}\left
(\frac 1{|\xi |}\right)\right]^2.\label{51}\end{equation}

This is just the case when the argument of the Jacobi and
Gegenbauer polynomials belongs to the interval $[-1,1]$.  Therefore
the asymptotics for $\mu_{nm}\gg 1$, $|n-m|\sim {\cal O}(1)$
is given by eq.  (26):

\begin{equation}W_n^m\approx\frac {\mbox{arctan}|\eta |}{|\eta |}\left
[J_{\frac {|n-m|}2}\left({\cal N}_{nm}\mbox{arctan}|\eta |\right)\right
]^2.\label{52}\end{equation}
If ${\cal N}_{nm}\mbox{arctan}|\eta |\gg 1$, the asymptotics of the
Bessel function yields

\begin{equation}W_n^m\approx\frac 2{\pi {\cal N}_{nm}|\eta |}\left
[\cos\left({\cal N}_{nm}\mbox{arctan}|\eta |-\frac {\pi}4(|n-m|+1
)\right)\right]^2.\label{53}\end{equation}

If $|n-m|\gg\mu_{nm}\sim {\cal O}(1)$, then eq. (28) leads to the asymptotics

\begin{equation}W_n^m\approx\sqrt {\frac 2{\pi\nu_{nm}}}\frac {|\eta
|^{|n-m|}}{2^{\mu_{nm}}\mu_{nm}!|\xi |^{|n-m|+1}}\left[H_{\mu_{nm}}\left
(\sqrt {\frac {|n-m|+1}{2|\xi |^2}}\right)\right]^2.\label{54}\end{equation}
For $|\xi |\sim {\cal O}(1)$ this formula can be simplified as follows,

\begin{equation}W_n^m\approx\sqrt {\frac 2{\pi\nu_{nm}}}\frac {\nu_{
nm}^{\mu_{nm}}|\eta |^{|n-m|}}{\mu_{nm}!|\xi |^{n+m+1}}.
\label{55}\end{equation}

Two other special cases correspond to small and large values of
the ``reflection coefficient $|\eta /\xi |^2$.  If $|\eta |\to 0$
(this limit corresponds
to an adiabatic change of the frequency $\omega (t)$, when $|\xi
|\to 1$), we get from eq.  (51)

\begin{equation}W_n^m\approx\frac {\nu_{nm}!}{\mu_{nm}!\left[\left
(\frac {|n-m|}2\right)!\right]^2}\left|\frac {\eta}2\right|^{|n-m
|}.\label{56}\end{equation}
For large values of $n$ and $m$ this expression coincides with the
limits of eqs.  (52) and (55) for $|\eta |\to 0$, if one takes into
account formula (22).

An interesting phenomenon is observed when $|\xi |\approx |\eta |
\gg 1$ (more
precisely, when $|\xi |\gg\nu_{nm}$).   Then for even values of $
n$ and $m$ we
get from eq.  (51)

\begin{equation}W_{2k}^{2l}\approx\frac 1{|\xi |}\frac {(2k-1)!!}{
(2k)!!}\frac {(2l-1)!!}{(2l)!!},\label{57}\end{equation}
whereas the transition probability between the neighbouring odd
levels is much less:

\begin{equation}W_{2k+1}^{2l+1}\approx\left[\frac {2\nu_{kl}+1}{|
\xi |}\right]^2W_{2k}^{2l}.\label{58}\end{equation}
For instance,

\[W_0^0=|\xi |^{-1},\qquad W_1^1=|\xi |^{-3},\qquad W_2^2\approx\frac
14|\xi |^{-1},\qquad W^3_3\approx\frac 94|\xi |^{-3},\ldots\]
So, for $|\xi |\gg\nu_{nm}$ the transition probabilities oscillate like the
photon distribution function in eq.  (47).  Such a coincidence is not
accidental, due to the relation [3]

\[|\xi |=\sqrt {{\cal E}+\frac 12},\]
${\cal E}$ being the normalized energy of quantum fluctuations in the final
Gaussian state which has originated from the initial ground state
with ${\cal E}_{\mbox{in}}=\frac 12$.  It is well known [3,21] that Gaussian
states with large
energy of fluctuations are highly squeezed, and a manifestation of
squeezing is just the oscillating structure of various distributions
related to the number of quanta.

\section{Acknowledgement}
One of us (V.I.M.)thanks INFN and University of Napoli "Federico II"
for the hospitality.

\newpage
\begin{center}
{\LARGE {\bf References}}\\
\end{center}

[1] I.A.Malkin, V.I.Man'ko, and D.A.Trifonov, J. Math. Phys. {\bf 14},
576 (1973).

[2] I.A.Malkin and V.I.Man'ko, {\em Dynamical Symmetries and Coherent
States of Quantum Systems\/} (Nauka, Moscow, 1979; in Russian).

[3] V.V.Dodonov and V.I.Man'ko, {\em Invariants and Evolution of
Nonstationary Quantum Systems}, Proceedings of Lebedev Physics
Institute, v.{\bf 183}, edited by M.A.Markov (English translation:  Nova
Science Publishers, Commack, N.Y.,  1989).

[4] M.Kauderer, J. Math. Phys. {\bf 34}, 4221 (1993).

[5] A.Vourdas and R.M.Weiner, Phys. Rev. {\bf 36}, 5866 (1987).

[6] G.Adam, Phys.Lett.A {\bf 171}, 66 (1992).

[7] I.A.Malkin, V.I.Man'ko, and D.A.Trifonov, Phys. Rev. D {\bf 2},
1371 (1970).

[8] V.V.Dodonov, I.A.Malkin, and V.I.Man'ko, Physica {\bf 59}, 241 (1972).

[9] V.V.Dodonov, V.I.Man'ko, and V.V.Semjonov,
Nuovo Cimento B {\bf 83}, 145 (1984).

[10] V.V.Dodonov, V.I.Man'ko, and I.N.Prokopenya, {\em Asymptotic Density
Matrix Elements of the Canonically Transformed Oscillator Hamiltonian
in the Fock Basis\/}, Preprint No. 30 (Lebedev Physics Institute, Moscow,
1985).

[11] A.Vourdas,  Phys.Rev.A  {\bf 34},  3466  (1986).

[12] {\em Bateman Manuscript Project: Higher Transcendental Functions},
edited by A.Erd\'elyi (McGraw-Hill, New York, 1953).

[13] G.Szeg\"o, {\em Orthogonal Polynomials\/} (American Math. Society,
New York, 1959).

[14] F.W.J.Olver, {\em Asymptotics and Special Functions\/} (Academic
Press, New York, 1974).

[15] {\em Handbook of Mathematical Functions\/}, edited by M.Abramowitz
and I.A.Stegun (National Bureau of Standards, Washington, D.C., 1964).

[16] S.Chaturvedi and V.Srinivasan, Phys.Rev.A {{\bf 40}}, 6095 (1989).

[17] V.V.Dodonov, O.V.Man'ko, and V.I.Man'ko, ``Photon distribution
for one-mode mixed light with generic gaussian Wigner function'',
University of Napoli preprint INFN-NA-IV-93/35, DSF-T-93/35(1993)
(submitted to Phys. Rev. A).

[18] P.Marian and T.A.Marian, Phys.Rev.A {\bf 47}, 4474 (1993).

[19] W.Schleich and J.A.Wheeler, J.Opt.Soc.Am.B {\bf 4}, 1715 (1987).

[20] V.V.Dodonov,  A.B.Klimov,  and  V.I.Man'ko,  Phys.Lett.A
{\bf 134}, 211 (1989).

[21] V.V.Dodonov,  A.B.Klimov,  and  V.I.Man'ko, in {\em Squeezed and
Correlated States of  Quantum Systems\/}, Proceedings of Lebedev Physics
Institute, v. {\bf 205}, edited by M.A.Markov (Nova Science, Commack, 1993).

\end{document}